\documentclass[
reprint,
amsmath,amssymb,
aps,
pra,
superscriptaddress,
]{revtex4-2}

\usepackage{tikz}
\usetikzlibrary{quantikz2}

\usepackage{multirow}
\usepackage{makecell}
\usepackage{amsmath}
\usepackage{mathrsfs}
\usepackage{physics} 
\usepackage{xcolor}
\usepackage{afterpage}
\usepackage{adjustbox}
\usepackage{rotating}
\usepackage[sort&compress]{natbib}
\usepackage{graphicx}
\usepackage{dcolumn}
\usepackage{bm}
\usepackage[colorlinks=true, urlcolor=blue, pdfborder={0 0 0}]{hyperref}
\usepackage{braket}
\hypersetup{
linkcolor=blue,
citecolor=blue
}

\begin{document}

\preprint{APS/123-QED}

\title{Quantum Jacobi-Davidson Method}

\author{Shaobo Zhang}
\email{shaozhang@student.unimelb.edu.au}
\affiliation{School of Physics, The University of Melbourne, Parkville, 3010, Australia}

\author{Akib Karim}
\affiliation{Data61, CSIRO, 3168, Clayton Australia}

\author{Harry M. Quiney}
\affiliation{School of Physics, The University of Melbourne, Parkville, 3010, Australia}

\author{Muhammad Usman}
\affiliation{School of Physics, The University of Melbourne, Parkville, 3010, Australia}
\affiliation{Data61, CSIRO, 3168, Clayton Australia}

\date{\today}

\begin{abstract}

Computing electronic structures of quantum systems is a key task underpinning many applications in photonics, solid-state physics, and quantum technologies. This task is typically performed through iterative algorithms to find the energy eigenstates of a Hamiltonian, which are usually computationally expensive and suffer from convergence issues. In this work, we develop and implement the Quantum Jacobi–Davidson (QJD) method and its quantum diagonalization variant, Sample-Based Quantum Jacobi–Davidson (SBQJD) method, and demonstrate their fast convergence for the ground-state energy estimation. We assess the intrinsic algorithmic performance of our methods through exact numerical simulations on a variety of quantum systems, including 8-qubit diagonally dominant matrices, 12-qubit one-dimensional Ising models, and 10-qubit water molecule (H$_2$O) Hamiltonian. Our results show that both QJD and SBQJD methods achieve significantly faster convergence and require fewer Pauli measurements when compared to the recently reported Quantum Davidson method, with SBQJD further benefiting from optimized reference state preparation. These findings establish the QJD framework as an efficient general-purpose subspace-based technique for solving quantum eigenvalue problems, providing a promising foundation for sparse Hamiltonian calculations on future fault-tolerant quantum hardware.

\end{abstract}

\maketitle


\section{\label{sec-Introduction}Introduction} 

Quantum computers are being developed around the world with the promise of solving computationally intensive problems which are currently intractable on classical computing platforms \cite{shor, quantum_finance, Max_West_adv_QML_1, gill2022quantum}. Amongst many potential applications of quantum computing, the calculation of electronic structures of many-body quantum systems is considered a central task \cite{cao2019quantum, mcardle2020quantum}. This is due to the fact that the computational cost and complexity of the electronic structure calculations grows exponentially with the size of the quantum systems, rendering them intractable on classical computers and positioning quantum computers as a promising alternative which fundamentally capture the quantum properties \cite{daley2022practical}. However, the capabilities of the current generation of quantum hardware are limited by noise and is therefore referred to as Noisy Intermediate-Scale Quantum (NISQ) devices \cite{preskill2018quantum}. In this NISQ era, quantum algorithms for computing the electronic structure properties are either based on variational approaches such as Variational Quantum Eigensolver (VQE) \cite{peruzzo2014variational, zhang2024full, poulin2014trotter, gokhale2019minimizing, vpyb-ynmz, PhysRevResearch.6.033223}, or subspace-based techniques which project problems onto a reduced subspace allowing computationally efficient hybrid quantum-classical formalisms \cite{takeshita2020increasing, mcclean2017hybrid, motta2020determining, chowdhury2024enhancing}.

While variational quantum circuits have been extensively studied on the NISQ devices to compute ground state and excited state properties \cite{peruzzo2014variational, cadi2024folded, nakanishi2019subspace, higgott2019variational, vpyb-ynmz, parrish2019quantum}, their performance is generally limited by challenges such as barren plateaus hindering classical optimization of parametrized gates \cite{larocca2025barren} and the exponentially growing dimensionality of the variational parameter space as a functions of the system size. Alternatively, the subspace-based methods, such as Quantum Subspace Expansion \cite{mcclean2017hybrid, takeshita2020increasing}, Quantum Lanczos \cite{motta2020determining}, and Quantum Davidson (QD) method \cite{tkachenko2024quantum}, have been shown to efficiently utilize quantum and classical resources to compute electronic structures of molecular systems. Notably, the recently proposed QD method \cite{tkachenko2024quantum}, inspired by the classical Davidson algorithm, combines quantum and classical computers to map the original Hamiltonian onto a Krylov subspace and iteratively expands this subspace until convergence is achieved. Such a method was shown to find the low-lying excited state energies of quantum systems with shallower circuit depths and fewer iterations than the Quantum Lanczos algorithm \cite{motta2020determining}. Our work further advances the field of subspace-based methods by proposing a new Quantum Jacobi-Davidson (QJD) method that demonstrates significantly faster convergence when directly compared with the QD method for a variety of investigated quantum systems.  

The proposed QJD method is inspired by the classical Jacobi–Davidson (JD) method \cite{sleijpen2000jacobi, sleijpen1998efficient} which extended the Davidson algorithm by imposing orthogonality constraints on the search directions of the subspace basis vectors, resulting in faster convergence for matrices that are not nearly diagonal or when effective preconditioning is available \cite{notay2004jacobi}, making JD method a more general framework for sparse matrix approximation. Despite its generality and favorable convergence properties, a quantum analogue of the JD method has not yet been developed. This motivated our work leading to the formulation of a quantum framework that can inherit the advantages of the classical JD method while enabling improved accuracy and efficient approximations on quantum–classical architectures.

In practice, hybrid quantum-classical methods rely on the availability of a reference state with a sufficiently large overlap with the true ground state. Recently, a hybrid quantum-classical approach, Sample-based Quantum Diagonalization (SQDiag) \cite{kanno2023quantum, robledo2025chemistry}, has been developed to approximate eigenvalues and eigenvectors of quantum systems by extracting the few most dominant basis vectors. This approach is particularly well suited to chemical systems, where the ground state is often dominated by only a few basis vectors. Moreover, it has been demonstrated that combining SQDiag with subspace projection methods enables the calculation of low-lying electronic states in quantum systems \cite{piccinelli2025quantum, barison2025quantum, duriez2025computing}. In addition to the QJD framework, we also combine QJD with the SQDiag method to further enhance convergence performance and denote the resulting hybrid algorithm as the Sample-Based Quantum Jacobi–Davidson (SBQJD) method. 

In order to demonstrate the working of our methods (QJD and SBQJD) and their comparative performance against the QD approach, we implement them for a variety of quantum systems which include 8-qubit diagonally dominant systems, 12-qubit one-dimensional Ising model, and 10-qubit water (H$_2$O) Hamiltonian. The diagonally dominant systems were synthetically created to show the intrinsic superiority of QJD and SBQJD methods as they are suitable for such systems. In practice, some quantum systems correspond to diagonally dominant Hamiltonian \cite{shayit2025numerically, windom2023iterative}, where QJD and SBQJD will naturally perform better than other computational techniques. On the other hand, one-dimensional Ising models and the water molecule are physically relevant systems which have been extensively studied for benchmarking quantum algorithms \cite{jones2024ground, nam2020ground, lolur2021benchmarking, tomesh2022supermarq}. In all cases, the results show that our methods converge much faster than the QD method and achieve accuracy well below the chemical accuracy for the quantum system. 

The rest of the paper is organized as follows. In Sec.~\ref{sec-Theory}, we present the theoretical foundations of this work: Sec.~\ref{subsec-JD} summaries the classical JD method, Sec.~\ref{subsec-QJD} introduces the QJD method, and Sec.~\ref{subsec-SQDiag} details the SBQJD method. In Sec.~\ref{sec-Experiment}, we evaluate the performance of our method for various quantum systems. Specifically, we discuss results for diagonal-dominant matrices (Sec.~\ref{subsec-dd-matrix}), a one-dimensional Ising model (Sec.~\ref{subsec-ising-model}), and water molecule (Sec.~\ref{subsec-water-molecule}). Finally, Sec.~\ref{sec-Conclusions} summarizes our research.




\section{\label{sec-Theory}Theory and  Methods}


\subsection{\label{subsec-JD}The Jacobi-Davidson Method}

The JD method was first proposed by Sleijpen and van der Vorst to iteratively expand a search subspace and solve the projected eigenvalue problem within that subspace to approximate a few eigenpairs of large symmetric matrices \cite{sleijpen2000jacobi}. For many systems of interest in quantum chemistry and solid-state physics, Hamiltonians are typically sparse and diagonally dominant \cite{shayit2025numerically, windom2023iterative}, making them naturally suited to the type of matrices for which the JD method is designed. Given such a Hamiltonian $H$, the JD method first applies the Rayleigh-Ritz procedure \cite{jia2001analysis}, which maps the original $H$ to a subspace $V$ spanned by a set of orthonormal basis vectors $\{\ket{v_1},\ket{v_2},...,\ket{v_n}\}$ to solve a subspace eigenvalue problem,
\begin{equation}
\label{eq-jd-ge}
\begin{aligned}
H'\ket{v_i'} &= E_i'\ket{v_i'}, 
\end{aligned}
\end{equation}
where
\begin{equation}
\label{eq-jd-ge1}
\begin{aligned}
H' &= V^\dagger HV.
\end{aligned}
\end{equation}
The corresponding Ritz value $E'_i$ and Ritz vector $\ket{rv} = V\ket{v'_i}$ form the Ritz pair $(E'_i, V\ket{v'_i})$, which is the current approximated solution of $H$. The residue vector is defined as $\ket{r} = (H-E'_iI)\ket{rv}$ in accordance with the Galerkin condition to be orthogonal to the entire subspace \cite{saad1980variations}. Convergence is assessed by checking whether the norm of the residual vector satisfies the threshold condition, $|\ket{r}|\le c$, where $c$ denotes the convergence tolerance.

The JD method then finds a vector $\ket{t}$, called the correction vector, to expand the subspace $V$ in an orthogonal direction to the current solution $\ket{rv}$ by approximating a correction equation
\begin{equation}
\label{eq-jd-correction}
\begin{aligned}
(I - \ket{rv}\bra{rv})(H-E'_iI)(I - \ket{rv}\bra{rv})\ket{t} = -\ket{r}. 
\end{aligned}
\end{equation}
The Gram-Schmidt process and normalization are applied to the correction vector $\ket{t}$ against $V$ due to the inevitable floating error accumulations from the linear equation approximation. The subspace $V$ is then expanded by adding $\ket{t}$ as $\ket{v_{n+1}}$, so $V = \{\ket{v_1},\ket{v_2},...,\ket{v_n},\ket{v_{n+1}}\}$. The expanded $V$ will then be substituted back into the Rayleigh-Ritz procedure for the next iteration.

The performance of the JD method heavily depends on how well $\ket{t}$ is approximated. For large systems, solving Eq.~\ref{eq-jd-correction} is often computationally expensive. We briefly review one preconditioner below that could reduce the cost and will be used in this work. Recall that the JD method searches an orthogonal direction of $\ket{rv}$, in other words, $\braket{t|rv} = 0$, so Eq.~\ref{eq-jd-correction} can further be written as 
\begin{equation}
\label{eq-jd-t}
\begin{aligned}
\ket{t} = \epsilon(H-E'_iI)^{-1}\ket{rv}-(H-E'_iI)^{-1}\ket{r}, 
\end{aligned}
\end{equation}
where
\begin{equation}
\label{eq-jd-t-epi}
\begin{aligned}
\epsilon = \frac{\bra{rv}(H-E'_iI)^{-1}\ket{r}}{\bra{rv}(H-E'_iI)^{-1}\ket{rv}}. 
\end{aligned}
\end{equation}
A feasible preconditioner $M \approx Diag(H)-E'_iI$ can be applied if $H$ is diagonally dominant, making the inverse calculation computationally efficient for classical hardware. In this case, Eq.~\ref{eq-jd-t} is approximated as
\begin{equation}
\label{eq-jd-precon}
\begin{aligned}
\ket{t} = \epsilon M^{-1}\ket{rv}-M^{-1}\ket{r}, 
\end{aligned}
\end{equation}
where
\begin{equation}
\label{eq-jd-precon-epi}
\begin{aligned}
\epsilon = \frac{\bra{rv}M^{-1}\ket{r}}{\bra{rv}M^{-1}\ket{rv}}. 
\end{aligned}
\end{equation}
For more detailed discussions of various preconditioners in the JD method, see Refs.~\cite{sleijpen2000jacobi, morgan2000preconditioning, sleijpen1998efficient}.

\subsection{\label{subsec-QJD}Quantum Jacobi-Davidson Method}

\begin{figure}
\begin{quantikz}
\lstick[1]{$q_a:\ket{0}^\otimes$}&\qwbundle{}&\gate[style={minimum width=0.85cm, minimum height=0.7cm}]{PR}&\gate[2]{SELECT}&\gate[style={minimum width=0.85cm, minimum height=0.7cm}]{PR^\dagger}&\meter{}\\
\lstick[1]{$q_d: \ket{rv}$}&\qwbundle{}&&&&
\end{quantikz}
\caption[Linear Combination of Unitaries]{
\label{fig-lcu} 
The quantum circuit for the linear combination of unitaries (LCU) method to obtain the normalized correction vector 
$\ket{t}/s = \bra{0}PR^\dagger \cdot SELECT \cdot PR\ket{0}\ket{rv} = A/s\ket{rv}$, where $A$ is a non-unitary matrix, $A = \sum_{i=1}^m \alpha_iU_i$, and $s$ is a normalization factor, $s = \sum_{i=1}^m |\alpha_i|$. $q_a$ and $q_d$ denote the ancilla and data qubits, respectively. The $PR$ gate constructs the state $PR\ket{0} = \sum_{i=1}^m\sqrt{|\alpha_i|/s}\ket{i}$ on the ancilla qubits, and the $ SELECT$ gate applies $SELECT\ket{i}\ket{rv} = \ket{i}U_i\ket{rv}$ on the data qubits.
}
\end{figure}
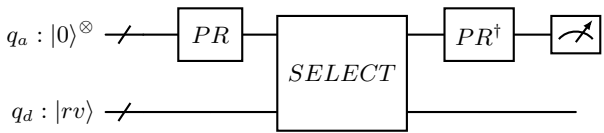

We propose the Quantum Jacobi-Davidson (QJD) method as a hybrid quantum–classical algorithm that employs quantum circuits to accelerate the computations of energy approximation and converges faster than the previously reported QD method \cite{tkachenko2024quantum}. Both Eq.~\ref{eq-jd-t} and Eq.~\ref{eq-jd-precon} have the form of $\ket{t} = A\ket{rv}$, where $A = \epsilon(H-E'_iI) - I$ and $A = \epsilon M^{-1} - M^{-1}(H-E'_iI)$, respectively. The matrix $A$ does not necessarily have to be unitary, but it can be decomposed into a linear combination of $m$ unitary operators, $A = \sum_{i=1}^m \alpha_iU_i$; therefore, the Linear Combination of Unitaries (LCU) is naturally chosen to approximate $\ket{t}$. We demonstrate the quantum circuit for preparing the quantum state $t$ using LCU in Fig.~\ref{fig-lcu}, where the circuit constructs the weighted state $A/s\ket{rv}$ on the data qubits when the ancilla qubits are measured in $0$ state, and the normalization factor $s = \sum_{i=1}^m |\alpha_i|$. The number of required ancilla qubits is $\left\lceil \log_2 m\right\rceil$. In the LCU method, the probability of measuring $0$ on the ancilla qubits is proportional to $1/s^2$, resulting in a scaling of $O(s^2\cdot polylog(s/p))$, where $p$ refers to the desired precision. To improve efficiency, amplitude amplification is often applied alongside LCU, providing a speedup of $O(s)$ and reducing the overall scaling to $O(s\cdot polylog(s/p))$ \cite{berry2015simulating}.

The correction equation of the QJD method can be interpreted as a Newton step for minimizing the Rayleigh quotient \cite{fokkema1998accelerated, zhou2006studies}; thus, it inherits the favorable local convergence properties of Newton-type algorithms. Although this equivalence has been previously reported \cite{sleijpen1995jacobi, fokkema1998accelerated}, we indendently derive it based on the Gâteaux derivative in Appendix.~\ref{sec-appen-QJD-Newton}, which provides a variational formulation that is directly compatible with quantum expectation value calculation, thereby completing and clarifying the theoretical foundation of the QJD method. In particular, once the reference state has sufficient overlap with the exact eigenstate, the QJD method exhibits at least a quadratic convergence in the eigenvalue error, in contrast to the linear convergence characteristic of first-order methods, such as the classical Davidson's method and QD method. Although iterations in both the QJD and QD methods with a full Hamiltonian preconditioner incur higher quantum resource costs due to the inverse operation, the quadratic reduction in the residual norm substantially decreases the total number of iterations required for convergence.

\begin{figure}[!h]
\begin{quantikz}
\lstick[1]{$\ket{rv}$}&\qwbundle{}&\gate[style={minimum width=0.85cm, minimum height=0.7cm}]{BS}&\meter{}
\end{quantikz}
\caption[Expectation Value Calculation]{
\label{fig-exp-calc} 
The quantum circuit for calculating the expectation value $\bra{rv}P_i\ket{rv}$, where $BS$ is a basis-changing gate constructed depending on the Pauli $X$, $Y$, and $Z$ gates present in $P_i$. 
}
\end{figure}
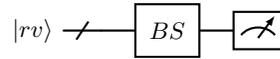

In contrast, the QD method with a residue vector preconditioner \cite{tkachenko2024quantum} can be interpreted as a first-order Krylov subspace approach, where the subspace expansion is driven by the residual vector rather than a Newton correction vector. As a result, the QD method avoids the explicit inversion of the shifted Hamiltonian and has a lower qubit-scaling per iteration, requiring only Hamiltonian application and overlap measurements. However, this reduction in circuit depth comes at the expense of linear convergence, leading to a substantially larger number of iterations for high-precision eigenvalue estimation. Consequently, when high accuracy is required or when a high-quality reference state is available, the QJD method offers a more favorable overall complexity trade-off despite its higher per-iteration cost, making it more suitable for fault-tolerant quantum devices.

\begin{figure}[!h]
\begin{quantikz}
\lstick[1]{$q_a:\ket{0}$}&\gate[style={minimum width=0.85cm, minimum height=0.7cm}]{H}&\ctrl{1}&\ctrl{1}&\ctrl{1}&\gate[style={minimum width=0.85cm, minimum height=0.7cm}]{H}&\meter{} \\
\lstick[1]{$q_d:\ket{0}^\otimes$}&\qwbundle{}&\gate[style={minimum width=0.85cm, minimum height=0.7cm}]{U_{r}}&\gate[style={minimum width=0.85cm, minimum height=0.7cm}]{P_i}&\gate[style={minimum width=0.85cm, minimum height=0.7cm}]{U^\dagger_{rv}}&&
\end{quantikz}
\caption[Hadamard Test]{
\label{fig-hadamard-test} 
The quantum circuit for calculating $\Re(\bra{rv}P_i\ket{r})$, where $q_a$ and $q_d$ denote the ancilla and data qubits, respectively. The states $\ket{rv}$ and $\ket{r}$ are constructed by $\ket{rv} = U_{rv}\ket{0}$ and $\ket{r} = U_{r}\ket{0}$, and $P_i$ represents a Pauli string.
}
\end{figure}
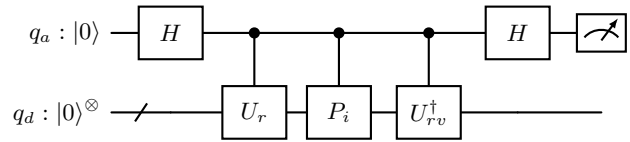

\begin{figure*}
\includegraphics[scale=0.8]{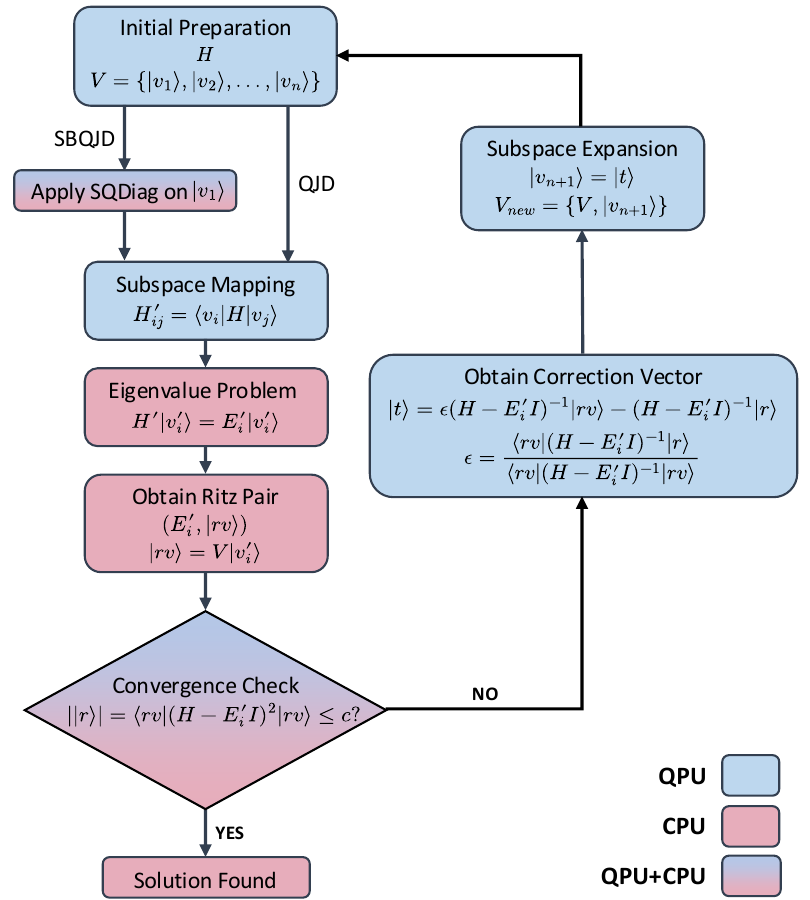}
\caption[Material structure]{
\label{fig1} 
The flowchart of the QJD and SBQJD methods. The blue, pink, and gradient-filled boxes indicate steps to be executed on the QPU, CPU, and in a hybrid QPU+CPU environment, respectively.
}
\end{figure*} 

For the calculation of $\epsilon$, Eq.~\ref{eq-jd-t-epi} can be further simplified to $\epsilon = 1/\bra{rv}(H-E'_iI)^{-1}\ket{rv}$, so only the expectation value evaluation is required for its denominator. The same approach can also be applied to compute the denominator in Eq.~\ref{eq-jd-precon-epi}. Since the inverse of the Hermitian matrices $(H-E'_iI)$ and $M$ are Hermitian, we shall use $B$ to denote a generic Hermitian matrix in the following paper. A Hermitian matrix $B$ can be decomposed into $B = \sum c_iP_i$, where $P_i$ is the Pauli string and $c_i$ is the corresponding coefficient. The expectation value of $B$ in terms of a given state $\ket{rv}$ is calculated as $\sum c_i\bra{rv}P_i\ket{rv}$, where each term $\bra{rv}P_i\ket{rv}$ can be obtained by a quantum circuit demonstrated in Fig.~\ref{fig-exp-calc}. The basis-changing gate ($BS$) is constructed as $H$, $S^\dagger H$, and $I$ with respect to the Pauli $X$, $Y$, and $Z$ gates shown in $P_i$, respectively. To calculate the overlap of two quantum states, $\bra{rv}B\ket{r} = \sum c_i\bra{rv}P_i\ket{r}$, as shown in the numerator in Eq.~\ref{eq-jd-precon-epi}, each term $\bra{rv}P_i\ket{r}$ can be estimated using a Hadamard test circuit. Since $\epsilon$ is defined as real in the JD method, we employ the circuit shown in Fig.~\ref{fig-hadamard-test} to calculate $\Re(\bra{rv}P_i\ket{r})$. The probability of measuring $0$ and $1$ on the ancilla qubit, $P(0)$ and $P(1)$, respectively, implying the overlap $\Re(\bra{rv}P_i\ket{r}) = 2P(0)-1 = 1-2P(1)$. The flowchart of QJD method is illustrated in Fig.~\ref{fig1}. The method starts from the initial preparation of $H$, and a reference state $\ket{v_1}$ as the first subspace basis vector, then iteratively expands the subspace by calculating the correction vector until it passes the convergence check. The blue, pink, and gradient-filled boxes indicate steps to be executed on the QPU, CPU, and in a hybrid QPU+CPU environment, respectively.

\begin{figure*}
\includegraphics[scale=0.2]{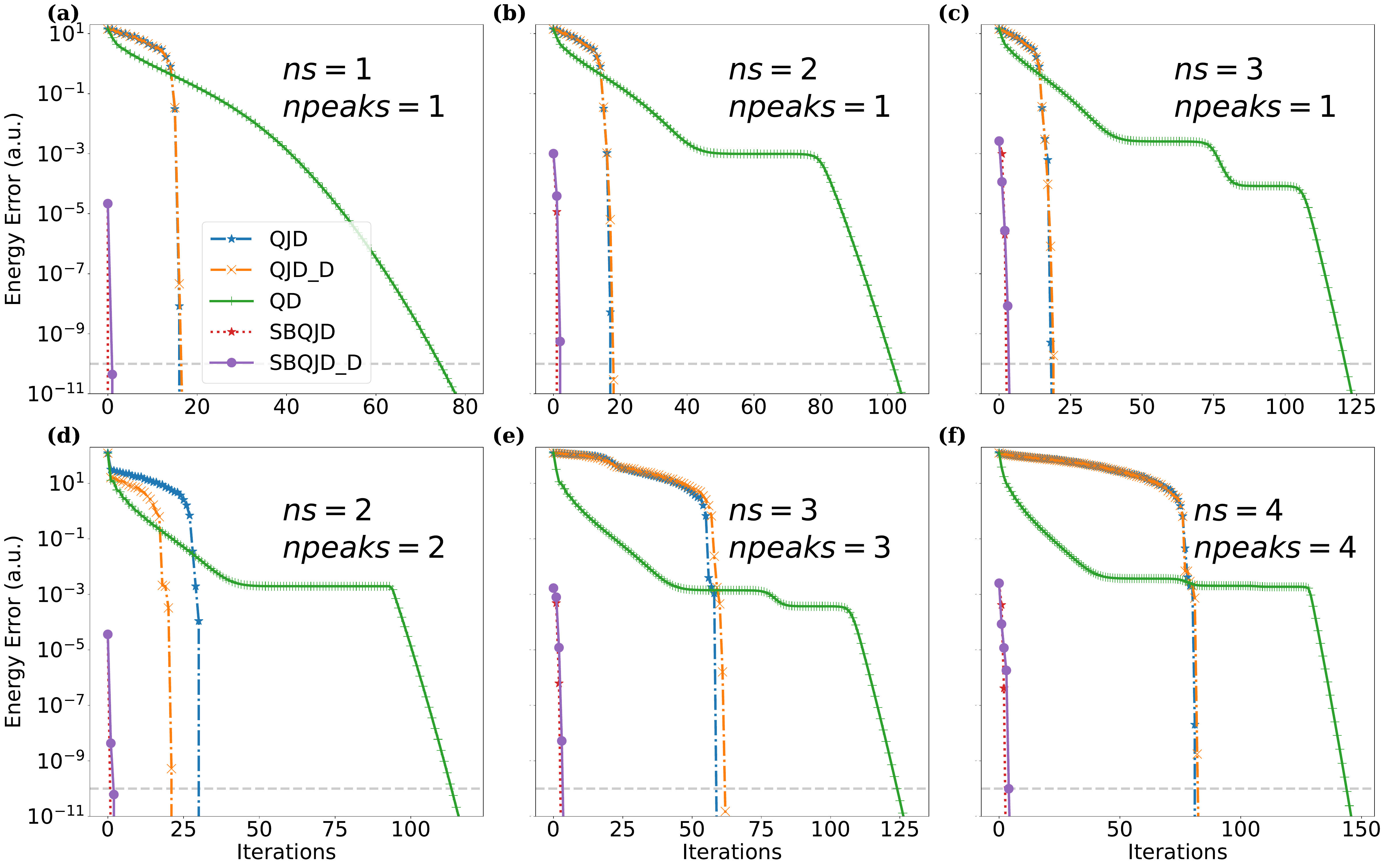}
\caption[Material structure]{
\label{fig2} 
The comparison of the QD, QJD, and SBQJD methods for approximating the ground state energy of six diagonal-dominant matrices. From left to right: the upper row shows a decreasing overlap between the reference state and the true ground state; the lower row demonstrates an increasing number of less dominant computational bases in the reference state. $ns$ refers to the number of smallest diagonal elements in the Hamiltonian, and $npeaks$ is defined as the number of peaks in the Gaussian distribution of the reference quantum state. The suffix $\_$D indicates the use of the diagonal Hamiltonian preconditioners for the calculation. The convergence tolerance is illustrated in a horizontal dashed grey line. The curves are truncated and displayed when the energy error is within the lower bound of the $y$-axis.
}
\end{figure*} 

\begin{figure}
\includegraphics[scale=0.21]{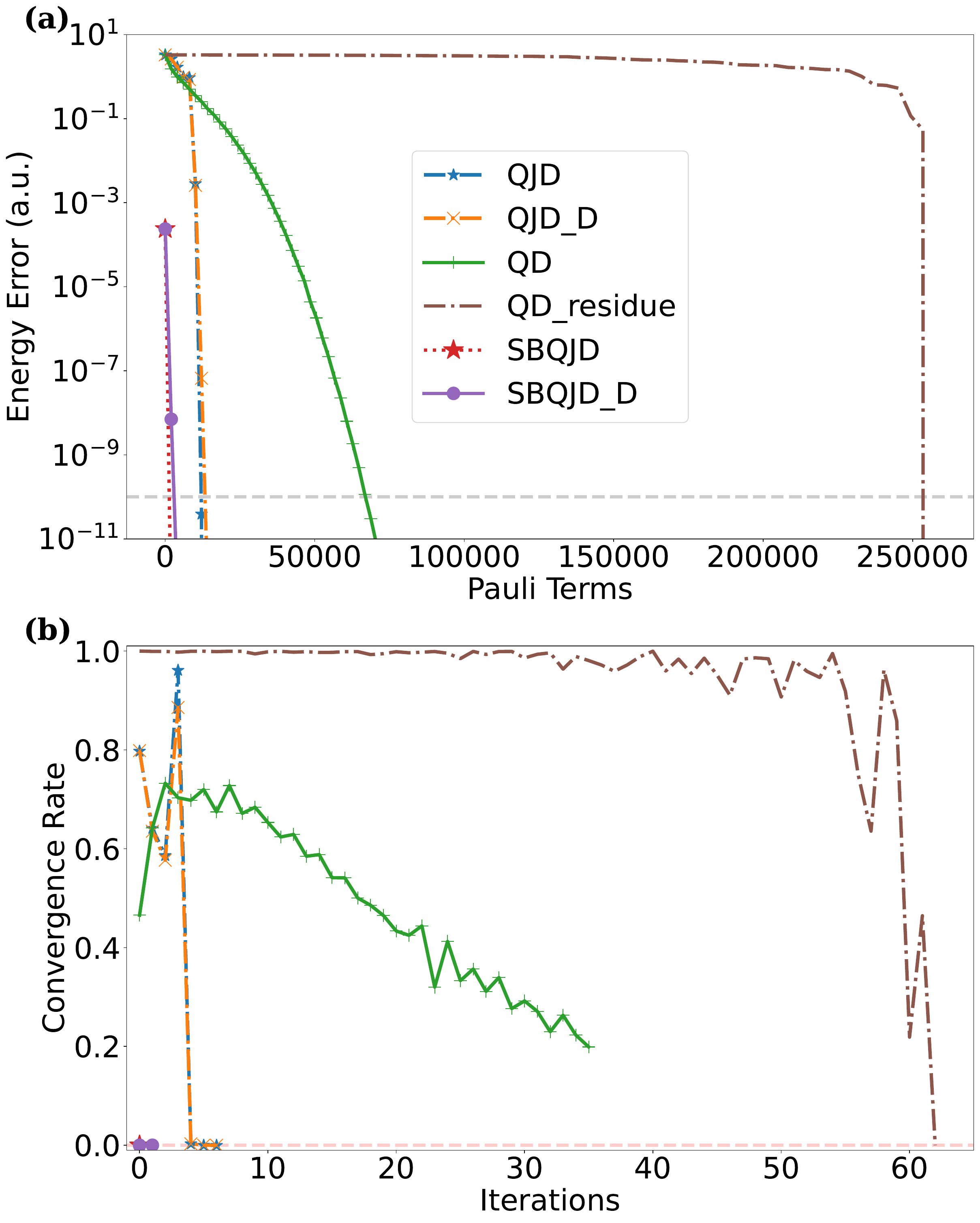}
\caption[Material structure]{
\label{fig3} 
The ground state energy calculations using the QD, QJD, and SBQJD methods, showing (a) the number of decomposed Pauli terms and (b) the convergence behavior in terms of the iteration number. The convergence tolerance is illustrated in a horizontal dashed grey line. The curves are truncated and displayed when the energy error is within the lower bound of the $y$-axis. QD$\_$residue denotes the use of the residue vector as the correction vector of the QD method \cite{tkachenko2024quantum}, and other notations in the legend are defined in the caption of Fig.~\ref{fig2}.
}
\end{figure} 

\subsection{\label{subsec-SQDiag}Sampled-Based Quantum Diagonalization}

The SQDiag method is a hybrid quantum–classical post-processing algorithm that approximates low-lying eigenstates of a Hamiltonian $H$ by diagonalizing the problem in a generated subspace \cite{kanno2023quantum, robledo2025chemistry}. Starting from the time-independent Schrodinger equation, $H\ket{\phi} = E\ket{\phi}$. The exponential growth of the Hilbert space with system size renders classical solutions intractable and brings significant challenges for quantum computers due to the measurement overhead associated with the Pauli-decomposed Hamiltonians $H$. Instead of measuring $H$ directly in the full Hilbert space of $\ket{\phi}$, the SQDiag method samples $\ket{\phi}$ on a quantum device and constructs a projected subspace $V_{SQDiag}$ spanned by $\{\ket{\phi'_0}, \ket{\phi'_1},...,\ket{\phi'_n}\}$. These states correspond to the $n$ most frequently observed computational-basis configurations in the measurement outcomes. The SQDiag method then maps the Hamiltonian $H$ onto the subspace $V_{SQDiag}$, $H_{SQDiag} = V^\dagger_{SQDiag} HV_{SQDiag}$, and the corresponding overlap matrix is an identity matrix due to the orthogonality between computational bases. Therefore, solving the reduced Hamiltonian problem, $H_{SQDiag}\ket{\phi_{SQDiag}} = E\ket{\phi_{SQDiag}}$, would obtain the Ritz pair $(E, V\ket{\phi_{SQDiag}})$, which is the approximated solution of $H$.

Both the QJD and SQDiag methods reduce computational cost by projecting the Hamiltonian onto a reduced subspace. In the QJD method, the subspace is spanned by a set of correction vectors $\{\ket{v_i}\}$. Projecting the Hamiltonian onto this subspace yields a reduced Hamiltonian with matrix elements $\bra{v_i}H\ket{v_j}$, where the diagonal and off-diagonal elements correspond to $i=j$ and $i\neq j$, respectively. As the system size grows, the matrix element $\bra{v_i}H\ket{v_j}$ can be approximated by $\bra{v^{SQDiag}_i}H\ket{v^{SQDiag}_j}$, where $\ket{v^{SQDiag}_i}$ and $\ket{v^{SQDiag}_j}$ are approximate representations of the correction vectors obtained using the SQDiag method. Therefore, this hybrid strategy could reduce the measurement cost and substantially accelerate the convergence of the QJD method. In this work, we employ the SQDiag method only in the first iteration of the QJD algorithm to enhance the overlap between the reference state and the true ground state. Extending SQDiag to all correction vectors is beyond the scope of the present study and is left for future work.

\begin{figure*}
\includegraphics[scale=0.21]{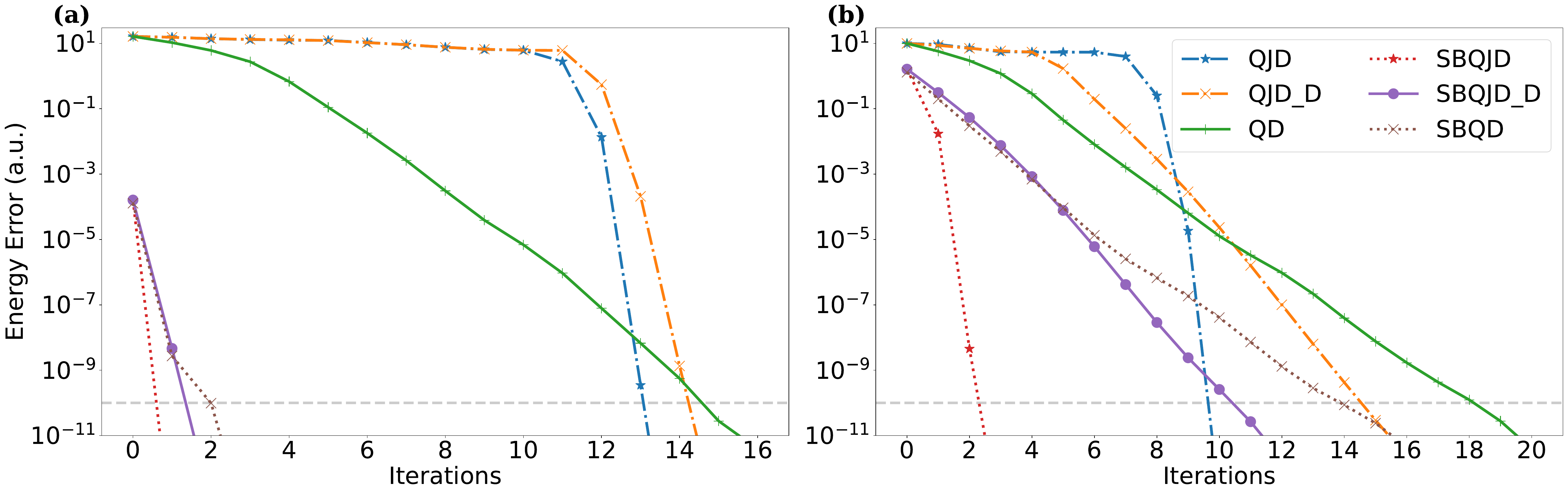}
\caption[Material structure]{
\label{fig4} 
(a) The ground state energy calculation of a one-dimensional Ising model with a diagonally dominant Hamiltonian. (b) The ground state energy calculation of a one-dimensional Ising model with a non-diagonally dominant Hamiltonian. The convergence tolerance is illustrated by a horizontal dashed grey line. The curves are truncated and displayed when the energy error is within the lower bound of the $y$-axis.The notations in the legend are defined in the caption of Fig.~\ref{fig2}.
}
\end{figure*}

\section{\label{sec-Experiment}Results and Discussions}

In this section, we present exact numerical simulations of the QJD and SBQJD methods to evaluate their intrinsic algorithmic performance. Since these approaches are not intended for direct implementation on NISQ hardware, analysis under noise is beyond the scope of this work. In particular, the algorithms rely on repeated coherent state preparation, accurate evaluation of Hamiltonian matrix elements, and iterative subspace updates, which collectively require circuit depths and error control beyond the capabilities of NISQ devices. We also note that the QD method \cite{tkachenko2024quantum} was likewise assessed exclusively through exact numerical simulations in its original proposal. We benchmark and compare the methods with the QD method on several representative systems, including diagonally dominant matrices (8 qubits), a one-dimensional Ising model (12 qubits), and a water molecule (10 qubits). The SBQJD method takes the three most dominant computational bases, $n=3$, for the calculation. The methods are considered converged when the energy error falls below $10^{-10}$. In all figures presenting our numerical results, the suffix indicates the preconditioner employed. Specifically, the suffix $\_$D represents the use of a diagonal preconditioner, in which only the diagonal elements of $H$ are used in the calculation, as described in Sec.~\ref{subsec-JD}; the suffix $\_$residue denotes the use of the residue vector as the preconditioner in the calculation. Methods without a suffix use the full Hamiltonian as the preconditioner.

\subsection{\label{subsec-dd-matrix}Diagonally Dominant Matrix}

The classical Davidson and Jacobi–Davidson methods were originally developed for quantum chemistry applications, where the Hamiltonians are typically sparse and diagonally dominant \cite{shayit2025numerically, windom2023iterative}. Notably, Hamiltonians based on nearest-neighbor tight-binding models represent such a class of systems and are widely used to compute electronic structure and properties of semiconductor materials. Motivated by this structure, we assess the performance of the QJD and SBQJD methods on a set of manually constructed diagonally dominant matrices of dimension 256 by 256, corresponding to an eight-qubit Hamiltonian, which we denote uniformly by $H$. The smallest eigenvalue of $H$ and its corresponding eigenvector are referred to as the ground-state energy and the ground state, respectively. The matrix is diagonally dominant if $|H_{i,i}|\geq \sum _{j\ne i}|H_{i,j}|$. Therefore, the off-diagonal elements of each matrix $H_{i,j}$ are randomly sampled from the interval $[0, 1/256]$ and fixed to be identical for all matrices considered in this scenario, while the diagonal elements of each matrix are initialized as $H_{i,i} = i$. 

We calculate the ground state energy of $H$ using the QD, QJD, and SBQJD methods, and demonstrate the corresponding experimental results in Fig.~\ref{fig2}. The $x$-axis denotes the iteration number, while the $y$-axis represents the energy error, which is defined as the difference between the exact energy and the computed energy. $ns$ refers to the number of the smallest diagonal elements of $H$. The reference state, $\ket{v_1}$, introduced in Sec.~\ref{subsec-JD}, is initialized as a quantum state whose amplitudes follow an approximately Gaussian distribution in the computational basis centered at the position of the smallest diagonal element of $H$. Since the ground state of a diagonally dominant Hamiltonian is strongly influenced by this minimum, such a choice ensures a substantial overlap between the reference state and the true ground state. Hence, we introduce $npeaks$ to demonstrate the number of peaks in the Gaussian distribution of the reference state. The grey horizontal dashed line illustrates the convergence tolerance.

We first examine the performance of our methods in the case where the reference state has a reduced overlap with the true ground state, and show the results in the upper row of Fig.~\ref{fig2}. Specifically, for Fig.~\ref{fig2}(a), $H_{1,1} = 1$ is the smallest diagonal element; for Fig.~\ref{fig2}(b), $H_{1,1} =H_{256,256}= 1$ are set to the smallest diagonal element; for Fig.~\ref{fig2}(c), $H_{1,1}=H_{128,128}=H_{256,256}=1$ are set to the smallest diagonal element. All the initial states are centered at the position of $H_{1,1}$. The QJD methods with both preconditioners take around 18 iterations to reach convergence for all three scenarios. The SBQJD method converge rapidly with both preconditioners, even as the initial overlap between the reference state and the true ground state decreases. The QD method requires more iterations as the overlap diminishes and exhibits additional plateaus in the subspace as $ns$ increases.

We then demonstrate the simulation results of the case where the peaks in the initial state matches the position of smallest diagonal elements of $H$ in Fig.~\ref{fig2}(d) ($H_{1,1} =H_{256,256}= 1$), Fig.~\ref{fig2}(e) ($H_{1,1}=H_{128,128}=H_{256,256}=1$), and Fig.~\ref{fig2}(f) ($H_{1,1}=H_{85,85}=H_{170,170}=H_{256,256}=1$), respectively. Similar to the case shown in the upper row of Fig.~\ref{fig2}, the two preconditioned SBQJD methods always converge rapidly. This behavior arises because the SQDiag procedure consistently identifies the three most significant computational basis states, thereby constructing reference states with larger overlap with the true ground state than that obtained from a Gaussian distributed initialization. Although the reference state has four peaks in Fig.~\ref{fig2}(f), capturing three of the four dominant computational basis states via the SQDiag method is more effective than including a larger number of nearby but less significant basis states. As $npeaks$ increases, the number of iterations required by the two preconditioned QJD methods becomes comparable and increases overall; however, it remains smaller than that required by the QD method, which is slowed by flatten parameter landscapes.

\begin{figure*}
\includegraphics[scale=0.26]{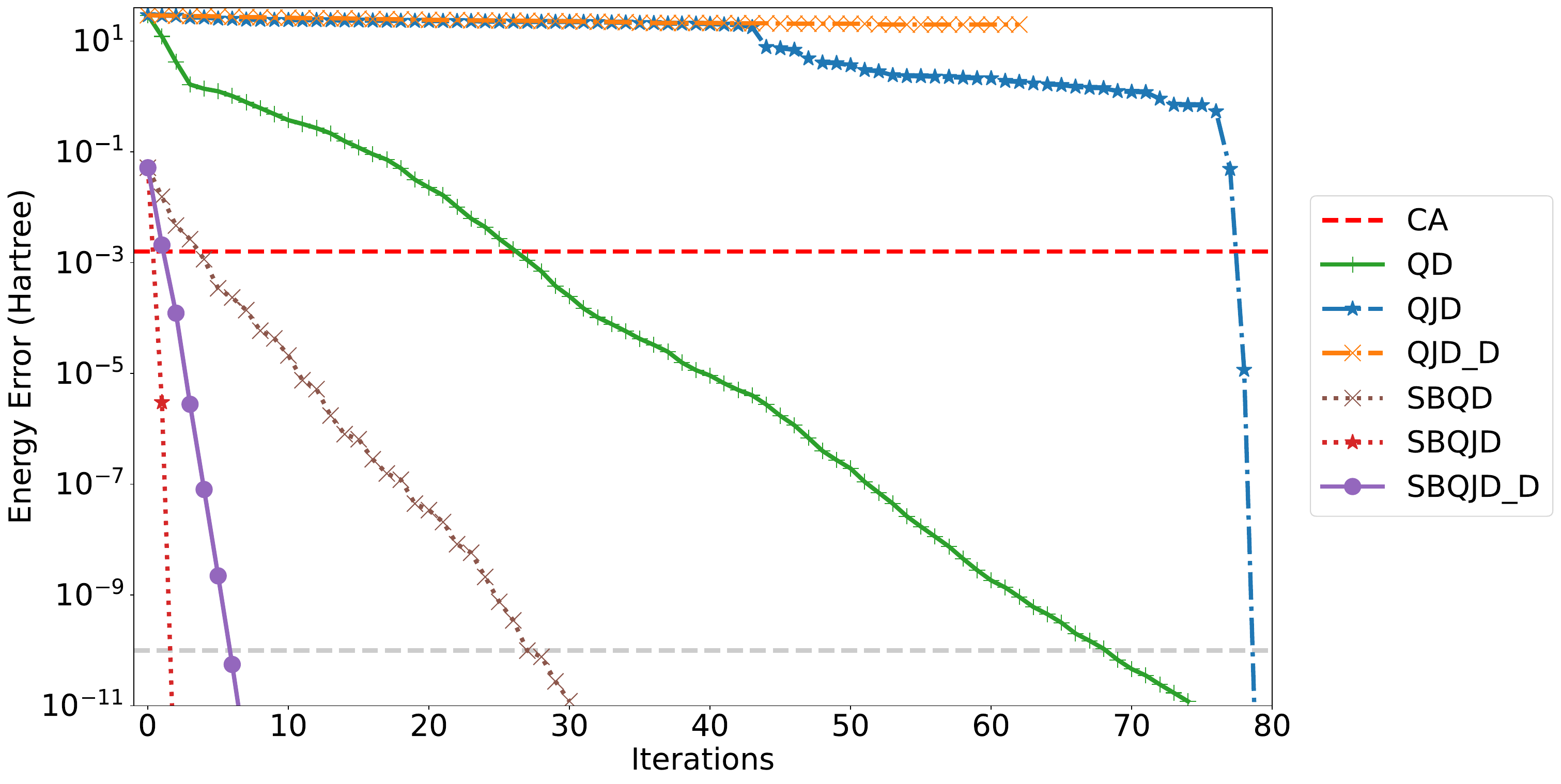}
\caption[Material structure]{
\label{fig5} 
The ground state energy calculation of a water molecule. The horizontal red and grey dashed lines show the chemical accuracy (CA) and convergence tolerance, respectively. The curves are truncated and displayed when the energy error is within the lower bound of the $y$-axis. SBQD denotes the combination of SQDiag and QD methods, and other notations in the legend are defined in the caption of Fig.~\ref{fig2}. 
}
\end{figure*} 

Furthermore, we present the performance of the QD, QJD, and SBQJD methods under the same scenario as Fig.~\ref{fig2} (a) in terms of decomposed Pauli terms in Fig.~\ref{fig3} (a). The QD$\_$residue refers to the preconditioner that takes the residue vector as the correction vector \cite{tkachenko2024quantum}. A generic $q$-qubit Hamiltonian contains $O(4^q)$ Pauli terms in its decomposition. By achieving a given target accuracy using substantially fewer measured Pauli terms, the QJD and SBQJD methods significantly reduce the measurement overhead, making them well-suited for execution on early fault-tolerant quantum devices. The associated convergence rate of each method at $i$th iteration is defined as $|E_i - E_{exact}|/|E_{i-1}-E_{exact}|$, and is shown in Fig.~\ref{fig3} (b). It remains at one when the method exhibits a flatten parameter space, and it is zero when the method finds the solution, as demonstrated by the QD$\_$residue and SBQJD methods, respectively.

\subsection{\label{subsec-ising-model}One-Dimensional Ising Model}

In this section, we consider a one-dimensional transverse-field Ising model to assess the performance of our methods. This model is paradigmatic for studying quantum phase transitions and is widely used as a benchmark for quantum simulations~\cite{dziarmaga2010dynamics, sachdev1999quantum, neira2025benchmarking, lin2021real}. The model is defined as 
\begin{equation}
\label{eq-ising-model}
\begin{aligned}
H = -J \sum_{i=1}^{n} \sigma_i^z \sigma_{i+1}^z \;-\; h \sum_{i=1}^{n} \sigma_i^z \;-\; g \sum_{i=1}^{n} \sigma_i^x, 
\end{aligned}
\end{equation}
where $\sigma^z$ and $\sigma^x$ are Pauli Z and X matrices. We select $\sigma_{n+1}^z = \sigma_1^z$ to satisfy the periodic boundary condition. Throughout this case, energies are expressed in arbitrary units, with the nearest-neighbor coupling energy $J=1.1$ setting the scale. $h$ is the strength of an external longitudinal field, and $g$ is the strength of the transverse field. Such a model becomes diagonally dominant when the magnitude of each diagonal matrix element exceeds the sum of the magnitudes of its corresponding off-diagonal elements. In practice, this condition can be satisfied by selecting $g$ that is sufficiently small relative to $J$ and $h$, or by increasing $J$ and $h$ such that the transverse-field-induced quantum effects are effectively suppressed. Importantly, only the relative ratios between $J$, $h$, and $g$ are physically relevant, as an overall rescaling of the Hamiltonian leaves the structure of the problem unchanged. In this work, we consider two variants of the 12-qubit one-dimensional Ising model. The first is chosen to be diagonally dominant, with parameters $J=1.1$, $h=0.9$, and $g=0.01$, while the second is non–diagonally dominant, with $J=1.1$, $h=0.9$, and $g=1$. The corresponding results are shown in Fig.~\ref{fig4}(a) and (b), respectively. To investigate how the performance changes when the reference state contains more weight on the dominant computational basis states ($n$), we apply the SQDiag method with $n=3$ in the first iteration of all three approaches, namely Sample-Based Quantum Davidson (SBQD) method, SBQJD, and SBQJD$\_$D.

For the Ising model with a diagonally dominant Hamiltonian [Fig.~\ref{fig4}(a)], the QJD method with both preconditioners exhibits an initially flat landscape in the parameter space, followed by a rapid reduction in the energy error. This accelerated convergence is consistent with the locally quadratic convergence characteristic of the Newton step, allowing QJD to reach the convergence tolerance in a similar number of iterations as the QD method.
By applying SQDiag to the first iteration, all the sample-based methods converge less than three iterations. For the Ising model with a non-diagonally dominant Hamiltonian [Fig.~\ref{fig4}(b)], the SBQJD method exhibits the fastest convergence, reaching convergence within three iterations. The QJD method surpasses the SBQJD$\_$D method and converges in 10 iterations. The QJD and SBQJD methods that employing a diagonal Hamiltonian as the preconditioner converge more slowly than those using the full Hamiltonian preconditioner; nevertheless, they still achieve convergence in fewer iterations than the QD and SBQD methods. These results indicate that QJD-based methods remain effective for more general Hamiltonians and that employing a resource-efficient preconditioner can still yield faster convergence than the QD method.

\renewcommand{\arraystretch}{1.2}
\begin{table*}
\caption{Summary of the number of iterations required to reach convergence for each method in the case of diagonally dominant matrices, the one-dimensional Ising model, and the water molecule.}
\label{tab-summary-QJD}
\centering
\begin{tabular}{|c|c|c|c|c|c|}
\hline
 &  & \textbf{QJD} & \textbf{SBQJD} & \textbf{QD} & \textbf{SBQD} \\
 &  & (our work) & (our work) & Ref.~\cite{tkachenko2024quantum} & (our work) \\
\hline
\multirow{6}{*}{\parbox{3.5cm}{\centering \textbf{Diagonally Dominant Matrix} (8 Qubits)}} 
 & $ns=1$, $npeaks=1$ & 16 & 1 & 77 &  \\
\cline{2-6}
 & $ns=2$, $npeaks=1$ & 17 & 2 & 106 &  \\
\cline{2-6}
 & $ns=3$, $npeaks=1$ & 16 & 3 & 124 &  \\
\cline{2-6}
 & $ns=2$, $npeaks=2$ & 29 & 2 & 114 &  \\
\cline{2-6}
 & $ns=3$, $npeaks=3$ & 61 & 3 & 124 &  \\
\cline{2-6}
 & $ns=4$, $npeaks=4$ & 76 & 3 & 146 &  \\
\hline
\multirow{2}{*}{\parbox{4.5cm}{\centering \textbf{One-Dimensional Ising Model} (12 Qubits)}} 
 & Diagonally Dominant & 14 & 1 & 15 & 2 \\
\cline{2-6}
 & Non-Diagonally Dominant & 10 & 3 & 19 & 14 \\
\hline
\textbf{Water Molecule} (10 Qubits) & & 78 & 2 & 68 & 27 \\
\hline
\end{tabular}
\end{table*}

\subsection{\label{subsec-water-molecule}Water Molecule}

In this section, we evaluate the performance of our methods for a realistic molecular system. Specifically, we consider a 10-qubit Hamiltonian of a water molecule detailed in Appendix.~\ref{sec-appen-water}, and show the simulation results in Fig.~\ref{fig5}. The reference state is constructed by centering it on the corresponding Hartree-Fock (HF) state $\ket{0011110011}$, with $10\%$ of the amplitude distributed symmetrically to neighboring configurations on either side. We note that, for SQDiag method, setting $n=1$ with this reference state recovers the HF state itself.

The QJD method with the diagonal Hamiltonian preconditioner fails to converge in this scenario. The apparent termination of the iteration suggests convergence to an incorrect state, for which the residual norm falls below the prescribed threshold despite the approximated eigenpair being inaccurate. The QJD method with the full Hamiltonian preconditioner takes 78 iterations to reach the convergence. During the first 43 iterations, its trajectory closely follows that of the QJD$\_$D method; it then identifies the correct search direction in parameter space and ultimately converges to the ground state. This behavior indicates that a computationally efficient preconditioner does not necessarily guarantee convergence to the correct solution. Interestingly, the QD method converges in approximately 70 iterations, making it faster than the QJD method given the prepared reference state. 

By applying SQDiag to the reference state, all three algorithms exhibit a significant improvement in performance. Both SBQJD and SBQJD$\_$D methods converge faster than the SBQD method, requiring two and seven iterations, respectively. In contrast, the SBQD method converges in approximately 27 iterations, which is still fewer than required by the original QD method. Therefore, for realistic systems, it is crucial that the dominant basis states are placed sufficient weight in the reference state to ensure reliable convergence to the correct solution, which can be achieved by applying the SQDiag method for the approximated state.


\section{\label{sec-Conclusions}Summary and Outlook}


In this work, we generalize the classical Jacobi–Davidson method to a quantum computing framework, which we refer to as the Quantum Jacobi–Davidson or QJD method, for calculating the ground state energy of quantum systems. We further introduce the SBQD and SBQJD methods by combining the SQDiag (as described in Sec.~\ref{subsec-SQDiag}) with the QD method of Ref. \cite{tkachenko2024quantum} and with our QJD method, respectively, to enhance the efficiency and accuracy of the energy approximation. We demonstrate the working of our methods by simulating a set of diagonally dominant matrices, one-dimensional Ising models, and water molecule. A summary of the simulation results of our methods is provided in Table.~\ref{tab-summary-QJD}, which shows that for all investigated quantum systems, our methods converge faster than the previously reported QD method. Interestingly, we find that when the reference state contains a large number of computational bases that have little overlap with the true ground state, the convergence of the QD method reaches the solution faster than the QJD method. However, when the SQDiag method is applied in the first iteration to retain only the dominant computational bases in the reference state, the QJD method exhibits the fastest energy error suppression speed and reaches convergence in substantially fewer iterations than the QD and SBQD methods. 

Although in this work our focus has been to find the ground state energies of the quantum systems, the subspace-based structure of the QJD framework makes it particularly well suited for targeting low-lying excited states, and will be an important direction for future work. Additionally, further developments may include efficient preconditioning strategies, adaptive subspace expansion schemes, and integration with state preparation methods such as SQDiag.  



In summary, we have demonstrated our methods using exact numerical simulations to assess the intrinsic algorithmic performance, which will be suited to implement on future fault-tolerant quantum hardware. Due to the requirements for the coherent state preparation and accurate Hamiltonian evaluation, QJD and SBQJD algorithms belong to the fault-tolerant quantum algorithm category (as is the case for the QD method \cite{tkachenko2024quantum} and other recently proposed quantum algorithms such as Refs. \cite{m6nc-ypl7, PhysRevApplied.23.054054, dqcnn}), which is beyond the scope of the current generation of NISQ devices. Our work has expanded the class of subspace-based quantum methods for the efficient computation of electronic structures in solid-state and molecular physics.



\begin{acknowledgments}

The research was supported by the University of Melbourne through the establishment of the IBM Quantum Network Hub at the University.

\end{acknowledgments}


\bibliography{main}

\onecolumngrid
\appendix

\section{\label{sec-appen-QJD-Newton}The Jacobi-Davidson Method and Newton's Method}


In this section, we will prove that the JD method is a projected Newton's method using the Gateaux derivative. Given a Hamiltonian $H$, $H\ket{v_i} = E_i\ket{v_i}$, where $E_i$ and $\ket{v_i}$ are the corresponding eigenvalue and eigenvector, $|\ket{v}| = 1$, we define a residue function 
\begin{equation}
\label{eq-residue-function}
\begin{aligned}
F(\ket{x}) = (H-\hat{E}I)\ket{x}
\end{aligned}
\end{equation}
such that $F(\ket{x}) = 0$ when $\ket{x}$ is a solution of $H$. $\hat{E}$ is the Rayleigh quotient, $\hat{E} = \hat{E}(H,\ket{x}) = \bra{x}H\ket{x}$, $|\ket{x}| = 1$, making $F(\ket{x})$ a non-linear eigenproblem. The JD method iteratively searches an orthogonal direction to the current solution; therefore, we can write the approximated solution $\ket{x_{n+1}} = \ket{x_n}+\epsilon\ket{x'}$, $\ket{x'}$ is the searching direction, where $\ket{x'}\perp\ket{x_n}$, and $\epsilon$ is a scaler. 

The Gateaux differential of the mapping $f: X\to Y$ in the direction $\Delta x$ in a Banach space is defined as 
\begin{equation}
\label{eq-gateaux}
\begin{aligned}
\delta_{\Delta x} f(x) = \lim_{\eta\to0} \frac{f(x+\eta\Delta x)-f(x)}{\eta}
= f'(x)\Delta x,
\end{aligned}
\end{equation}
where $f'(x)$ is a linear operator called the Gateaux derivative of $f(x)$ at $x$. Substituting $\hat{E}$ into $f$, $\ket{x}$ into $x$, and $\ket{x}+\epsilon\ket{x'}$ into $x+\eta\Delta x$ in eq.~\ref{eq-gateaux}, we obtain the Gateaux differential of $\hat{E}$: 
\begin{equation}
\label{eq-gateaux-rayleigh}
\begin{aligned}
\delta _{\ket{x'}}\hat{E} = &\lim_{\epsilon\to0} \frac{\hat{E}(H, \ket{x}+\epsilon\ket{x'})-\hat{E}(H,\ket{x})}{\epsilon} \\
=& 
\lim_{\epsilon\to0}\frac{(\bra{x}+\epsilon\bra{x'})H(\ket{x}+\epsilon\ket{x'}) - \bra{x}H\ket{x}}{\epsilon} \\
=&
\lim_{\epsilon\to0}\frac{\bra{x}H\ket{x}+\epsilon(\bra{x'}H\ket{x}+\bra{x}H\ket{x'})+\epsilon^2\bra{x'}H\ket{x'}-\bra{x}H\ket{x}}{\epsilon} \\
=&
\lim_{\epsilon\to0}\frac{\epsilon(\bra{x'}H\ket{x}+\bra{x}H\ket{x'})+\epsilon^2\bra{x'}H\ket{x'}}{\epsilon} \\
=&
\bra{x'}H\ket{x}+\bra{x}H\ket{x'} \\
=&
\begin{cases}
2\bra{x'}H\ket{x}, &\text{for real $H$}; \\
\Re{2\bra{x'}H\ket{x}}, &\text{for complex $H$}.
\end{cases}
\end{aligned}
\end{equation}
When $\ket{x}$ is an eigenvector of $H$, $F(\ket{x})=0$, so $H\ket{x} = \hat{E}\ket{x}$, 
\begin{equation}
\label{eq-gateaux-x'hx1}
\begin{aligned}
&\bra{x'}H\ket{x} \\
=& \bra{x'}\hat{E}\ket{x} \\
=& \hat{E}\braket{x'|x};
\end{aligned}
\end{equation}
when $\ket{x}$ is an approximated eigenvector of $H$, $F(\ket{x})\neq0$,
\begin{equation}
\label{eq-gateaux-x'hx2}
\begin{aligned}
&\bra{x'}H\ket{x} \\
=& \bra{x'}(\hat{E}I+H-\hat{E}I)\ket{x} \\
=& \hat{E}\braket{x'|x} + \bra{x'}(H-\hat{E}I)\ket{x} \\
=& \hat{E}\braket{x'|x} + \bra{x'}F(\ket{x}).
\end{aligned}
\end{equation}
When the correction equation in the JD method is solved exactly, since the searching direction is always perpendicular to the current approximated solution, $\braket{x'|x} = 0$, Eq.~\ref{eq-gateaux-x'hx1} is always 0, and the residue $F(\ket{x})\to0$ when $\ket{x}$ is close to the exact solution, making Eq.~\ref{eq-gateaux-x'hx2} $\to0$; when the correction equation is solved approximately, as $\ket{x}$ is iteratively approaching the exact eigenvector, the residue $F(\ket{x})\to0$, making $\ket{x'} \to 0$, as a result, $\text{Eq.~\ref{eq-gateaux-x'hx1}} \approx \text{Eq.~\ref{eq-gateaux-x'hx2}} \to 0$. Accordingly, $\delta _{\ket{x'}}\hat{E} \approx 0$ when the approximated eigenvector $\ket{x}$ has a non-trivial overlap with the exact eigenvector. Similarly, the Gateaux differential of $F(\ket{x})$ is 
\begin{equation}
\label{eq-gateaux-JD}
\begin{aligned}
\delta _{\ket{x'}}F = &\lim_{\epsilon\to0} \frac{F(\ket{x}+\epsilon\ket{x'})-F(\ket{x})}{\epsilon} \\
=& 
\lim_{\epsilon\to0}\frac{
[H-\hat{E}(H, \ket{x}+\epsilon\ket{x'})](\ket{x}+\epsilon\ket{x'}) - [H-\hat{E}(H,\ket{x})]\ket{x}
}{\epsilon} \\
=&
\lim_{\epsilon\to0}\frac{
H\ket{x}+\epsilon H\ket{x'}-[\bra{x}H\ket{x}+\epsilon(\bra{x'}H\ket{x}+\bra{x}H\ket{x'})+\epsilon^2\bra{x'}H\ket{x'}](\ket{x}+\epsilon\ket{x'}) - [H-\hat{E}(H,\ket{x})]\ket{x}}{\epsilon} \\
=&
\lim_{\epsilon\to0}\frac{
H\ket{x}+\epsilon H\ket{x'}-[\hat{E}(H,\ket{x})+\epsilon\delta _{\ket{x'}}\hat{E}+\epsilon^2\hat{E}(H,\ket{x'})](\ket{x}+\epsilon\ket{x'}) - [H-\hat{E}(H,\ket{x})]\ket{x}
}{\epsilon} \\
=&
\lim_{\epsilon\to0}\frac{
\epsilon(H-\hat{E}(H,\ket{x})\ket{x'}-\delta_{\ket{x'}}\hat{E}\ket{x})+O(\epsilon^2)+O(\epsilon^3)
}{\epsilon} \\
=&
[H-\hat{E}(H,\ket{x})]\ket{x'}-\delta_{\ket{x'}}\hat{E}\ket{x} \\
\approx&
[H-\hat{E}(H,\ket{x})]\ket{x'},
\end{aligned}
\end{equation}
the Gateaux derivative $F'(\ket{x}) = H-\hat{E}(H,\ket{x})$ on $\ket{x}$.

Newton’s method iteratively refines an approximated solution to a non-linear equation by linearizing the function at each step and updating the estimate using the inverse of its derivative. Given a function $g(m)$, it is defined as 
\begin{equation}
\label{eq-newton}
\begin{aligned}
-g(m_n) = g'(m_n)\Delta m,
\end{aligned}
\end{equation}
where $g'(m)$ is the derivative of $g(m)$ and $\Delta m = m_{n+1}-m_n$. Since the right side of both Eq.~\ref{eq-gateaux} and Eq.~\ref{eq-newton} expresses the derivative that is a linear map acting on a perturbation direction, we substitute Eq.~\ref{eq-residue-function} and Eq.~\ref{eq-gateaux-JD} into Eq.~\ref{eq-newton}, then obtain
\begin{equation}
\label{eq-jd-newton1}
\begin{aligned}
-F(\ket{x}) = [H-\hat{E}(H,\ket{x})]\ket{x'}.
\end{aligned}
\end{equation}
Let $P = (I-\ket{x}\bra{x})$ be a projection operator that projects any vector onto the subspace orthogonal to $\ket{x}$, $P\ket{x} = 0$ and $P\ket{x'} = \ket{x'}$, Eq.~\ref{eq-jd-newton1} becomes
\begin{equation}
\label{eq-jd-newton2}
\begin{aligned}
-PF(\ket{x}) &= P[H-\hat{E}(H,\ket{x})]\ket{x'} \\
-F(\ket{x}) &= P[H-\hat{E}(H,\ket{x})]\ket{x'} \\
-F(\ket{x}) &= P[H-\hat{E}(H,\ket{x})]P\ket{x'}.
\end{aligned}
\end{equation}
By substituting $F(x)$ with $\ket{r}$ and $\ket{x'}$ with $\ket{t}$, we obtain the correction equation of the JD method introduced in Section.~\ref{subsec-JD},
\begin{equation}
\label{eq-jd-newton3}
\begin{aligned}
-\ket{r} = (I-\ket{x}\bra{x})[H-\hat{E}(H,\ket{x})](I-\ket{x}\bra{x})\ket{t},
\end{aligned}
\end{equation}
where $(\hat{E}(H,\ket{x}), \ket{x})$ is a Ritz pair forming an approximated solution of $H$, and $\ket{t}$ is a searching direction perpendicular to the approximated eigenvector $\ket{x}$. Hence, we prove that the correction equation of the JD method is a projected Newton step.

\section{\label{sec-appen-water}Water Hamiltonian}

The water Hamiltonian was generated using PySCF \cite{pyscf2018}. The molecule used fixed experimental ground state co-ordinates with oxygen atom at the centre at $[0,0,0]$ Angstroms; the two hydrogens atoms at $[0.758602, 0.0, 0.504284]$ and $[0.758602, 0.0, -0.504284]$ Angstroms. 

Restricted Hartree-Fock was performed with the STO-3G basis set for the singlet ground state to find the molecular orbitals. The one and two-electron integrals were calculated and transformed into the molecular orbital basis to give the Hamiltonian. We used Jordan-Wigner mapping and qubit tapering~\cite{bravyi2017} to obtain a ten-qubit Hamiltonian.

\end{document}